\documentclass[11pt]{article}

\usepackage{graphicx}
\usepackage{endfloat}
\usepackage{amsfonts}
\usepackage{mychicago}
\usepackage{subfigure}
\usepackage{genetics_manu_style}

\usepackage{amsfonts}
\usepackage{amssymb}
\usepackage{listings}
\usepackage{hyperref}
\usepackage{amsthm}
\usepackage{latexsym}
\usepackage{xspace}
\usepackage{graphicx}
\usepackage{amsmath}
\usepackage{epsfig}
\usepackage{mathrsfs}
\usepackage{url}
\newtheorem{ex}{Example}

\newcommand{\bx}{\boldsymbol{x}}

\newcommand{\bm}{\boldsymbol{m}}
\newcommand{\by}{\boldsymbol{y}}

\newcommand{\bg}{\boldsymbol{g}}

\newcommand{\be}{\boldsymbol{e}}
\newcommand{\bzero}{\boldsymbol{0}}

\usepackage{times}
\renewcommand{\CorrespondingAddress}{Plant Breeding \& Genetics \\
    Cornell University \\ 
    Ithaca, NY 14850\\ 
    
    (419) 308 7085 (ph.)  \\
    
    \texttt{da346@cornell.edu} \vfill}
\renewcommand{\RunningHead}{Locally Epistatic Models}
\renewcommand{\CorrespondingAuthor}{Deniz Akdemir}
\renewcommand{\KeyWords}{Genomic Selection, Epistatis, Mixed Models}

\title{Genomic Prediction of Quantitative Traits using Sparse and Locally Epistatic Models}
\author{Deniz Akdemir \\ Department of Plant Breeding \& Genetics\\ 
  Cornell University\\ Ithaca, NY}

\begin{document}

\maketitle

\begin{abstract}
In plant and animal breeding studies a distinction is made between the genetic value (additive + epistatic genetic effects) and the breeding value (additive genetic effects) of an individual since it is expected that some of the epistatic genetic effects will be lost due to recombination. In this paper, we argue that the breeder can take advantage of some of the epistatic marker effects in regions of low recombination. The models introduced here aim to estimate local epistatic line heritability by using the genetic map information and combine the local additive and epistatic effects. To this end, we have used semi-parametric mixed models with multiple local genomic relationship matrices with hierarchical designs and lasso post-processing for sparsity in the final model. Our models produce good predictive performance along with good explanatory information.   
%\begin{keywords}
%Genomic selection, Genome wide association, Plant / animal breeding, Mixed model, Multiple kernel learning, Heritability
%\end{keywords}
\end{abstract}

\section{Introduction}

Selection in animal or plant breeding is usually based on estimates of genetic breeding values (GEBV) obtained with semi-parametric mixed models (SPMM). In these mixed models genetic information in the form of a pedigree or markers are used to construct an additive kernel matrix that describes the similarity of line specific additive genetic effects. These models have been successfully used for predicting the breeding values in plants and animals. The studies show that using similarities calculated from sufficient genome wide marker information almost always lead to better prediction models for the breeding values compared to the pedigree based models. In both simulation studies and in empirical studies of dairy cattle, mice and in bi-parental populations of maize, barley and \emph{Arabidopsis} marker based SPMM GEBVs have been quite accurate.

A SPMM for the $n\times 1$ response vector $\by$ is expressed as 
\begin{equation}\label{eq:spmm} \by=X\beta+Z\bg+\be \end{equation} where $X$ is the $n\times p$ design matrix for the fixed effects, $\beta$ is a $p\times 1$ vector of fixed effects coefficients, $Z$ is the $n\times q$ design matrix for the random effects; the random effects $(\bg',\be')'$ are assumed to follow a multivariate normal distribution with mean $\bzero$ and covariance \[ \left( \begin{array}{cc}
\sigma^2_g K  & \bzero  \\
\bzero & \sigma^2_e I_n \end{array} \right)\] where $K$ is  a $q\times q$ kernel matrix.

The similarity of the kernel based SPMM's and reproducing kernel Hilbert spaces (RKHS) regression models has been stressed recently (\cite{gianola2008reproducing}). In fact, this connection was previously recognized by \cite{kimeldorf1970correspondence}, \cite{harville1983discussion}, \cite{robinson1991blup} and \cite{speed1991blup}. RKHS regression models use an implicit or explicit mapping of the input data into a high dimensional feature space defined by a kernel function. This is often referred to as the ''kernel trick'' (\cite{scholkopflearning}).

A kernel function, $k(.,.)$ maps a pair of input points $\bx$ and $\bx'$ into real numbers. It is by definition symmetric ($k(\bx,\bx')=k(\bx',\bx)$) and non-negative. Given the inputs for the $n$ individuals we can compute a kernel matrix $K$ whose entries are $K_{ij}=k(\bx_i,\bx_j).$ The linear kernel function is given by $k(\bx; \by) = \bx'\by.$ The polynomial kernel function is given by $k(\bx; \by) =(\bx'\by+ c)^d$ for $c$ and  $d$ $\in$ $R.$  Finally, the Gaussian kernel function is given by $k(\bx; \by) = \frac{1}{\sqrt{2\pi h}}exp(-(\bx'-\by)'(\bx'-\by)/2h)$ where $h>0.$ Taylor expansions of these kernel functions reveal that each of these kernels correspond to a different feature map. 

RKHS regression extends SPMM's by allowing a wide variety of kernel matrices, not necessarily additive in the input variables, calculated using a variety of kernel functions. The common choices for kernel functions are the linear , polynomial, Gaussian kernel functions, though many other options are available.

For the marker based SPMM's, a genetic kernel matrix calculated using a linear kernel matrix incorporates only additive effects of markers. A genetic kernel matrix based on the polynomial kernel of order $k$ incorporates all of the one to $k$ order monomials of markers in an additive fashion. The Gaussian kernel function allows us to implicitly incorporate the additive and complex epistatic effects of the markers.

Simulation studies and results from empirical experiments show that the prediction accuracies of models with Gaussian are usually higher than the models with linear kernel. However, it is not possible to know how much of the increase in accuracy can be transferred into subsequent generations because some of the predicted epistatic effects that will be lost by recombination. This issue touches the difference between the commercial value of a line which is defined as the overall genetic effect (additive+epistatic) and the breeding value which is the potential for being a good parent (additive) and it can be argued that linear kernel model estimates the breeding value whereas the Gaussian kernel estimates the genetic value. 

In this article, we argue that the breeder can take advantage of some of the epistatic marker effects in regions of low recombination. We will refer to a  set of markers in a linkage group as a locality and will form a prediction model that combines the genomewide additive marker and  genomewide local epistatic genetic effects in an additive fashion. Since the epistatic effects that are incorporated in the model are local there is little chance that these effects will disappear with recombination.

The final models we propose in this paper can be viewed as SPMMs with semi-supervised kernel matrices that are obtained as weighted sum of functions of many local kernels. The major aim of this article is to measure and incorporate additive and local epistatic genetic contributions since we believe that the local epistatic effects are relevant to the breeder. The local heritability models in this article can be adjusted so that genetic contribution of the whole genome, the chromosomes, or local regions can be obtained. 

In most genome wide association studies (GWAS) the focus is on estimating the effects of individual markers and lower level interactions. However, in the genomic era, the number of SNP markers can easily reach millions and the methods used in GWAS for large samples become computationally exhaustive.  The local kernel approach developed in this article remedies this problem by reducing the number of hypothesis by focusing on regions and testing the nested hypothesis in an hierarchy. 

Another argument for why we would like to focus on short segments of the genome as distinct structures comes from the ''building blocks'' hypothesis in the evolutionary theory. The schema theorem of Holland \cite{holland1975adaptation} predicts that a complex system which uses evolutionary mechanisms such as fitness, recombination and mutation tend to generate short and well fit and specialized structures, these basic structures serve as building blocks. For example, when the alleles associated to an important fitness trait are scattered all around the genome the favourable effects can easily be lost just by independent segregation, therefore inversions that clump these alleles together physically would be strongly selected for. 

Finally, the sum of the "building blocks" approach we propose in this paper are parsimonious since only a few genomic regions are utilized in the final model and usually more accurate than their linear kernel model or Gaussian kernel model counterparts.   In addition, importance scores for genomic regions are obtained as a by product.

The remaining of the paper is organized as follows: In the next section, after briefly reviewing some multiple kernel approaches from the statistics and machine learning literature, we introduce our model which is more suitable to use in the context of traditional SPMMs. We discuss the issues of model set-up, parameter estimation, hypothesis testing here in. In Section 3, we will illustrate our model with four benchmark data sets.  We finally conclude with our remarks. 

\section{Multiple Kernel Models with Mapped Markers}
\subsection{Multiple kernel learning}
In recent years, several methods have been proposed to combine multiple kernel matrices instead of using a single one. These kernel matrices may correspond to using different notions of similarity or may be using information coming from multiple sources. 

A good review and taxonomy of multiple kernel learning algorithms in the machine learning literature can be found in \cite{gonen2011multiple}. Some related literature worth noting include \cite{hartley1967maximum} and more recent \cite{bach2004multiple} and \cite{sonnenburg2006large}.

Multiple kernel learning methods use multiple kernels by combining them into a single one via a combination function. The most commonly used combination function is linear. Given kernels $K_1, K_2,\ldots ,K_p,$ a linear kernel is of the form \[K=\eta_1K_1+\eta_2 K_2+\ldots+\eta_p K_p.\] The components of $K$ are usually input variables from different sources or different kernels calculated from same input variables. The kernel $K$ can also include interaction components like $K_i\odot K_j,$  $K_i\otimes K_j,$ or perhaps $-(K_i-K_j)\odot (K_j-K_i).$ For example, if $K_E$ is the environment kernel matrix and $K_G$ is the genetic kernel matrix, then a component $K_E\odot K_G$ can be used to capture the gene by environment interaction effects.

The kernel weights $\eta_1, \eta_2, \ldots, \eta_p$ are usually assumed to be positive and this corresponds to calculating a kernel in the combined feature spaces of the individual kernels. Therefore, once given the kernels, multiple kernel learning boils down to estimating the kernel weights. 

\subsection{A Locally Epistatic Genomic Model for Genomic Association, Prediction}

Our model building approach has three stages:
\begin{enumerate}

\item Subsets of the genome: Divide the marker set into $k$ subsets.\\
\item Local genetic values (GEBV's): Use the training data to obtain a model to estimate the local genetic values $\hat{g}_{j}(\bm)$ for each genome region $j=1,2,\ldots, k.$ \\
\item Post-processing: Combine the local GEBVs using an additive model fitted in the training data set.
\end{enumerate}
In the remaining of this section we will describe each step in more detail. 

\subsubsection{Locally epistatic kernels from mapped marker data}

In order to obtain $k$ kernels for a marker data, we will need  $k$ possibly nested or overlapping subsets of the marker set. As we will be noted in the conclusions section, these subsets can be obtained using any annotation of the markers. However since our aim is to capture the additive + locally epistatic genetic effects in a model, we will concentrate only on the possibly nested and overlapping regions of the genome. A genomic region is defined as set of markers in a linkage group.  

\begin{figure}[h]
\centering
\includegraphics*[width=.5\textwidth]{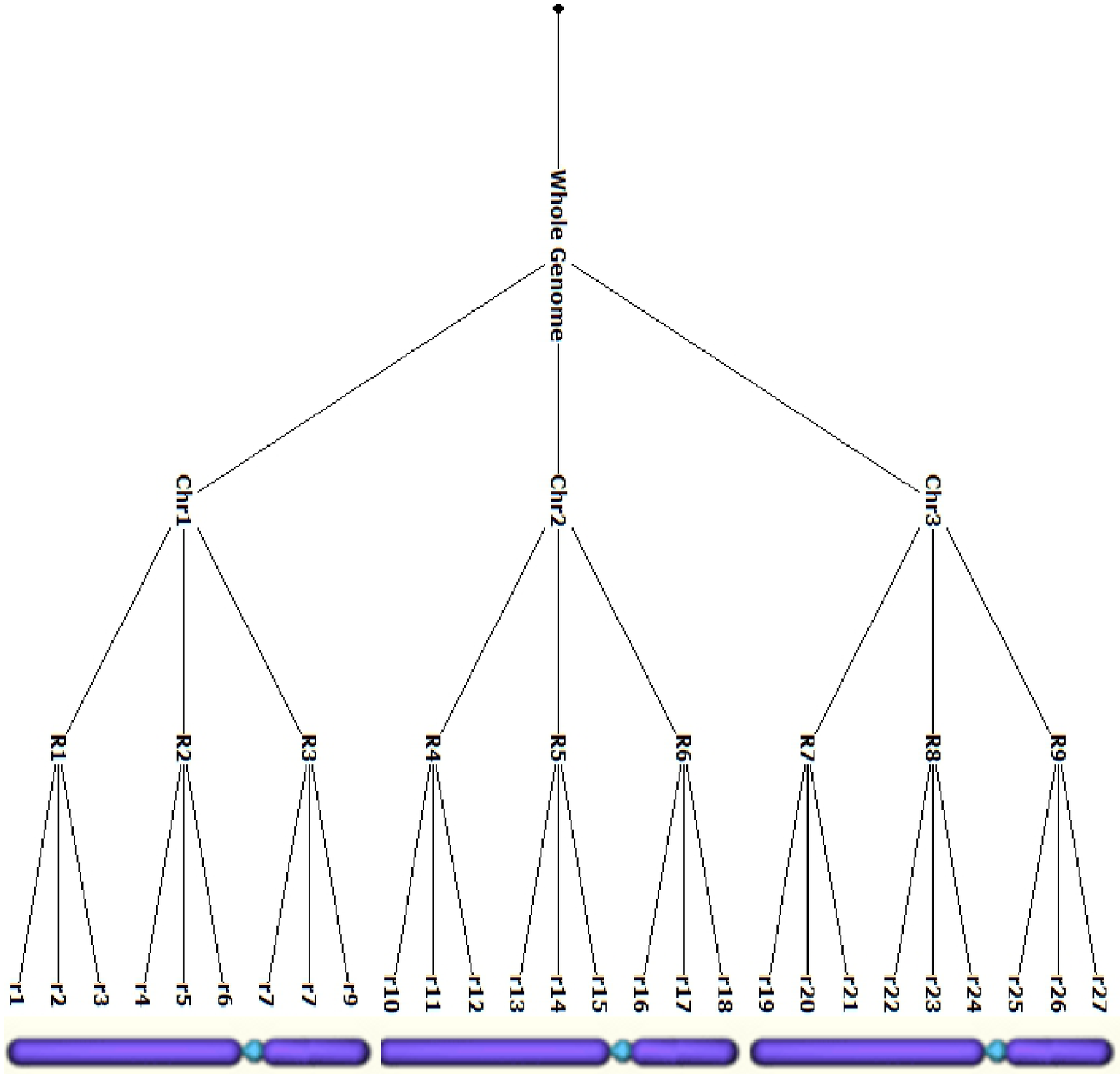}
\caption[Hierarchical set up]{An hypothetical hierarchical set up for an organism with 3 chromosomes.}
\label{fig:fighthier}
\end{figure}

Although it is possible to define genomic regions in an informed fashion, in our illustrations we will accomplish this task hierarchically as illustrated for an hypothetical organism with $3$ chromosomes in Figure \ref{fig:fighthier}. In this figure, at the root of the hierarchy we have the whole genome, second level of the hierarchy divides the genome into chromosomes, at the third level each chromosome is further divided into subregions, and so on.     

The hierarchical set up allows to define nested regions from coarse to fine, the regions are divided into subregions and this is repeated to a desired detail level. An advantage to using an hierachical set up is the availability of hierarchical testing procedures which have nice cost / power properties (\cite{blanchard2005hierarchical}). Multiple testing procedures where coarse to fine hypotheses are tested sequentially  have been proposed to control the family wise error rate or false discovery rate (\cite{reiner2003identifying}, \cite{meinshausen2008hierarchical}).  These procedures can be used along the ''keep rejecting until first acceptance'' scheme to test  hypotheses in an hierarchy. 

It is also useful to know that the a partitioning of the markers can be described by a factor variable. We will assume that the several partitionings of the markers which define $k$ possibly nested or overlapping regions is given and this is denoted by $P.$ 
%%%%

\subsubsection{Multiple kernel SPMMs}

Although some multiple kernel approaches use fixed weights for combining kernels, in most cases the weight parameters need to be learned from the training data. Some principled techniques used to estimate these parameters include likelihood based approaches in the mixed modeling framework like Fisher scoring algorithm or variance least squares approach though these approaches are more suitable to cases where only a few kernels are being used. \cite{qiu2009framework} propose two simple heuristics to select kernel weights in regression problems where the weights of kernels are either proportional to the correlations between the estimates from single kernel models and the response or they are proportional to the alignments of the kernels with the outer product of the response variable. 

The above methods fail to give satisfactory solutions in high dimensional settings, i.e, when the number of kernels is large or when the dimension of the kernels is large. This is mainly due to the difficulty in finding optimal parameter values. In the remaining of this section, we develop a model which is more suitable for use in high dimensional settings.

To obtain the  local genetic values one possible approach is to use a SPMM with multiple kernels in the form of 
\begin{equation}\label{eq:spmmmk} \by=X\beta+Z\bg_1 +Z\bg_2+\ldots+Z\bg_k+\be \end{equation} where $\bg_j\sim N_{q_k}(\bzero, \sigma^2_{g_j}K_j)$ for $j=1,2,\ldots,k,$ $\be\sim N_ {n}(0,\sigma^2_e I)$ and $\bg_1,$ $\ldots,$ $\bg_k,$ $\be$ are mutually independent.   

Another model incorporates the marginal variance contribution for each kernel matrix.  For this we use the following SPMM:
\begin{equation}\label{eq:spmmmk2} \by=X\beta+Z\bg_j+Z\bg_{-j}+\be_j \end{equation} where $\bg_j\sim N_{q_k}(\bzero, \sigma^2_{g_j}K_j)$ for $j=1,2,\ldots,k.$ $\bg_{-j}\sim N_{q_k}(\bzero, \sigma^2_{g_{-j}}K_{-j})$ is the random effect corresponding to the input components other than the ones in group $j$ and $K_{-j}$ stands for the kernel matrix obtained from markers not in group $j.$ $\be_j\sim N_{n}(0,\sigma^2_{e_j} I)$ and   $\bg_j,$ $\bg_{-j},$ $\be_j$ are mutually independent.    

A simpler approach is to use a separate SPMM for each kernel. Let $\hat{\sigma}^2_{g_j}$ and $\hat{\sigma}^2_{e_j}$ be the estimated variance components from the SPMM model in (\ref{eq:spmm}) with kernel $K=K_j.$ The markers corresponding to the random effect $\bg_{-j}$ which mainly accounts for the sample structure can now be incorporated as a fixed effect via their  principal components, say the matrix of first few principal components of the markers not in group $j$ is denoted by the matrix $PC_{-j}$. The model is written as  \begin{equation}\label{eq:spmmmk3} \by=X\beta+Z PC_{-j}\tau_{-j}+Z\bg_j+\be_j \end{equation} where $\tau_{-j}$ is considered as a fixed effect, $\bg_j\sim N_{q_k}(\bzero, \sigma^2_{g_j}K_j)$ for $j=1,2,\ldots,k,$ $\be_j\sim N_{n}(0,\sigma^2_{e_j} I)$ and   $\bg_j,$ $\be_j$ are independent. In the remaining of this paper we will combine the fixed effect terms in one as $X^*\beta^*$ for notational ease.  

The estimates of parameters for models in (\ref{eq:spmm}), (\ref{eq:spmmmk}) and  (\ref{eq:spmmmk2}) can be by maximizing the likelihood or the restricted (or residual, or reduced) maximum likelihood (REML). There are very efficient algorithms devised for estimating the parameters of the single kernel model in  (\ref{eq:spmmmk3}) and therefore this model will be our preferred model for the remaing of this paper. Estimating the parameters of Model (\ref{eq:spmmmk}) gets very difficult with large number of kernels and with large sample sizes, the single kernel or the marginal kernel models are more suitable in such cases.  

In (\cite{crainiceanu2003likelihood}), an approximate test is developed for testing the significance of an individual random effect $\bg_j$ for model \ref{eq:spmmmk3}. To deal with the inflation of the error probabilities due to testing $k$ hypothesis in the hierarchical set up,  we will use Meinshausen's hierarchical testing procedure (\cite{meinshausen2008hierarchical}) that controls the family wise error by adjusting the significance levels of single tests in the hierarchy. The procedure starts testing the root node $H_0$ at level $\alpha.$ When a parent hypothesis is rejected one continues with testing all the child nodes of that parent. The significance level to be used at each node $H$ is adjusted by a factor proportional to the number of variables in that node: \[\alpha_H=\alpha\frac{|H|}{|H_0|}\] where $|.|$ denotes the cardinality of a set. This means that larger penalty is incurred at finer levels. 

Once the fixed effects and the variance parameters of the model \ref{eq:spmmmk3} are estimated for the $j$th region, the expected value of the genetic effects (EBLUP) specific to region $j$ can be estimated by \[\hat{g}_{j}=\hat{\sigma}^2_{g_j}K_jZ'(\hat{\sigma}^2_{g_j}ZK_jZ'
+\hat{\sigma}^2_{e_j}I)^{-1}(\by-X^*\hat{\beta^*})\] for $j=1,2,\ldots, k$ for $j=1,2,\ldots,k.$

\subsubsection{Post-processing}
Let $\bx$ be the $p$ vector of fixed effects and $\bm$ be the vector of markers partitioned into $k$ regions.  Also, let $\hat{g}_{j}(\bm)$  denote the standardized EBLUPs of random effect components that correspond to the $k$ local kernels for regions $j=1,2,\ldots, k$ and individual with markers $\bm.$ Consider a final prediction model in the following form:
\begin{equation}f(\bx,\bm; \beta, \alpha)=\beta_0+\sum_{j=1}^{k}\alpha_{j}\hat{g}_{j}(\bm)+\sum_{j=k+1}^{k+p}\beta_j x_j.\label{eq:additivemodel}\end{equation}

Estimate the model coefficients using the following loss function \begin{equation}(\hat{\beta}, \hat{\alpha})=\underset{(\beta, \alpha)}{\operatorname{argmin}} \sum_{i=1}^N(y_i-(\beta_0+\sum_{j=1}^{k}\alpha_{j}\hat{g}_{j}(\bm_i)+\sum_{j=k+1}^{k+p}\beta_j x_{ji}))^2+\lambda \sum_{j=1}^k|\alpha_j|.\end{equation}
$\lambda>0$ is the shrinkage operator, larger values of $\lambda$ decreases the number of models included in the final prediction model.  

When $k$ is large compared to the sample size $N,$ we should use the following loss function with the elastic-net penalty  \small
\begin{equation}(\hat{\beta}, \hat{\alpha})=\underset{(\beta, \alpha)}{\operatorname{argmin}} \sum_{i=1}^N(y_i-(\beta_0+\sum_{j=1}^{k}\alpha_{j}\hat{g}_{j}(\bm_i)+\sum_{j=k+1}^{k+p}\beta_j x_{ji}))^2+\lambda_1 \sum_{j=1}^k|\alpha_j|+\lambda_2 \sum_{j=1}^k(\alpha_j)^2\end{equation} \normalsize to allow for more than $N$ non zero coefficients in the final estimation model. $\lambda_1,\lambda_2>0$ are the shrinkage operators.

In our examples, we have used $\widehat{G(\bm)}\widehat{\alpha}$ as the estimated genotypic value for an individual with markers $\bm$ where $\widehat{G(\bm)}=(\hat{g}(\bm)_1,\hat{g}(\bm)_2,\ldots,\hat{g}(\bm)_k).$ Since $\widehat{G(m)}$ has standardized columns, $|\widehat{\alpha}|$ can be used as importance scores for the regions in the model. 
\subsubsection{Hyper-parameters of the model}

While fitting the model in \ref{eq:additivemodel} we need to decide on the values of a number of hyper-parameters. Apart from the model set-up that involve the definition of genomic regions, these parameters are the kernel parameters and the parameters related to the elastic-net used in the post-processing step. 

Some standard ways of performing hyper-parameter optimization include grid search or random search, these methods need to be guided by some performance metric, typically measured by cross-validation on the training set or evaluation on a held-out validation set. In this paper we did not explore in detail how the model hyper-parameters should be chosen. Instead, we have used predetermined values throughout our illustrations and reported their results.  It is possible that the accuracies of the models can be improved by setting the hyperparameters in a more data dependent and informed approach. 

In our opinion, the hyperparameter choice for the multiple kernel model should reflect the available resources and the aims of the researcher. For instance, the number of regions that we can define depends on the number of markers, and a more detailed analysis might only be suitable when the number of markers and the number of examples in the training dataset are large. The hyperparameters of the shrinage estimators in the postprocessing step allows us to control the sparcity of the model. These parameters can be optimized for generalization performance but their value can also be influenced by the amount of sparcity desired in the model.  The multiple kernel models provide the user with the flexibility of models of various detail and sparcity.

\section{Illustrations}
 
In this section, for four datasets which represent a variety of situations, we will compare the locally epistatic model with its counterpoints linear and Gaussian kernel SPMMs. In particular, we report the correlation between the phenotypic values in the test data and the corresponding estimated genotypic values from our models. The lasso importance scores obtained for the genome regions are also provided.

\begin{ex}(Wheat Data)
This data was downloaded from The Triticeae Toolbox (\url{triticeaetoolbox.org}). 3735 markers on seven chromosomes for 337 elite wheat lines (SW-AMPanel) were available for the analysis. The traits (flowering date (FD), heading date (HD), physiological maturity date (MD), plant height (PH), yield (YD), waxiness (WX)) were obtained in four trials during years 2012 and 2013. The data is also available as a supplementary material.  We have sampled $90\%$ lines for training the models and we have used the rest of the lines to  evaluate the fit of our models. The whole genome was divided in a similar fashion as displayed in Figure \ref{fig:fighthier} with a depth of two. The accuracies of the multiple kernel model  compared with the linear and Gaussian kernel SPMMs and the mean  genome-wide importance scores for regions used in our multiple kernel model over 30 replications of the experiment are summarized for different choices of number of splits in Figures 2-9.   In addition, the importance scores from the multiple kernel model are used to cluster the traits and the resulting similarity of the traits are described by the dendograms in Figures 6-9. 

\begin{figure}[h]
\centering
\begin{minipage}[b]{0.45\linewidth}
\includegraphics*[angle=630,width=1\textwidth]{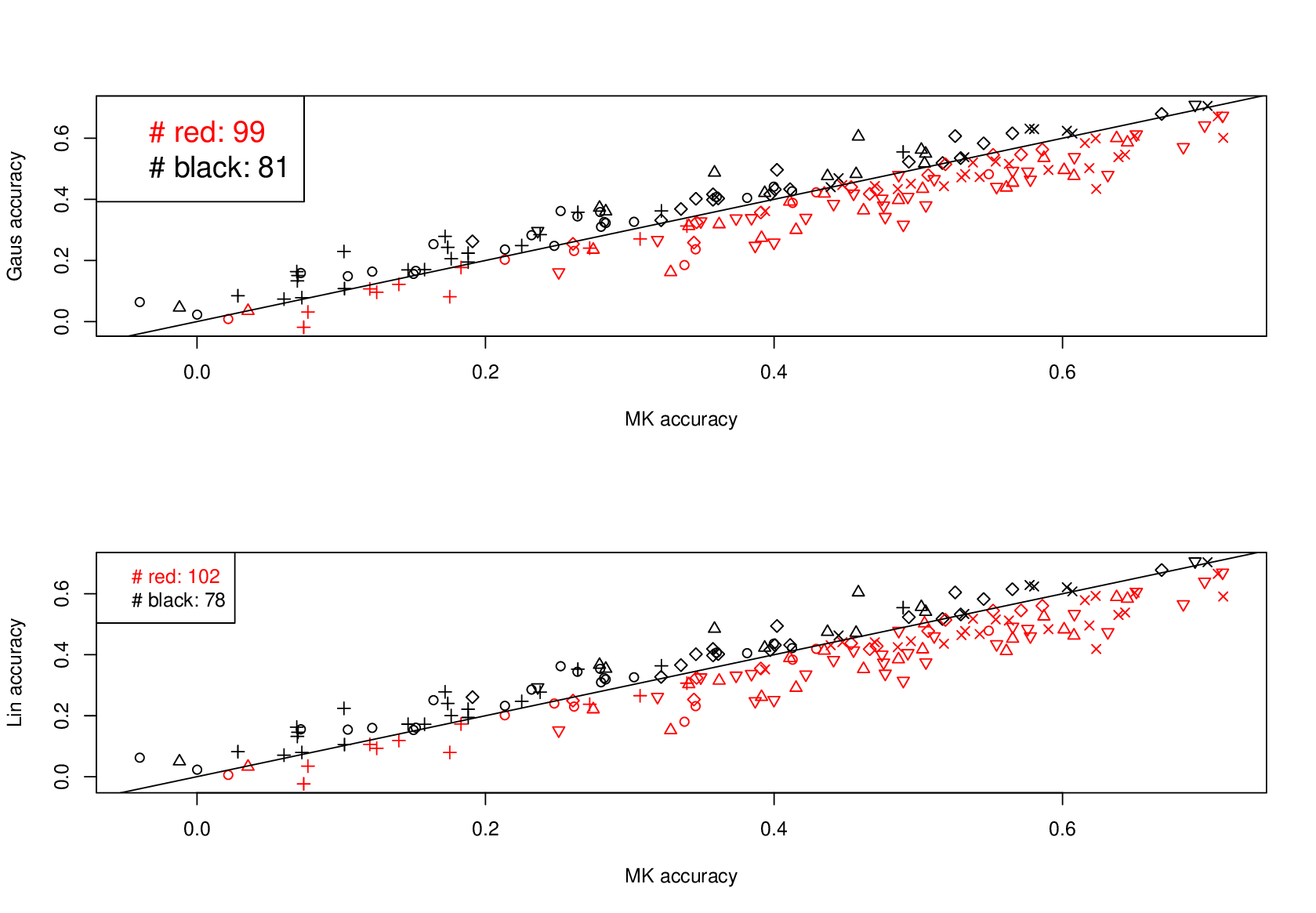}
\caption[Wheat Data: Accuracies 1]{Wheat Data: Accuracies of multiple kernel (MK) model (2 regions per chromosome) compared to Gaussian (Gaus) kernel model for the six traits. Red dots below the line correspond to cases MK model is more accurate than Gaus model. }
\label{fig:minipage39ex1}
\end{minipage}
\quad
\begin{minipage}[b]{0.45\linewidth}
\includegraphics*[angle=630,width=1\textwidth]{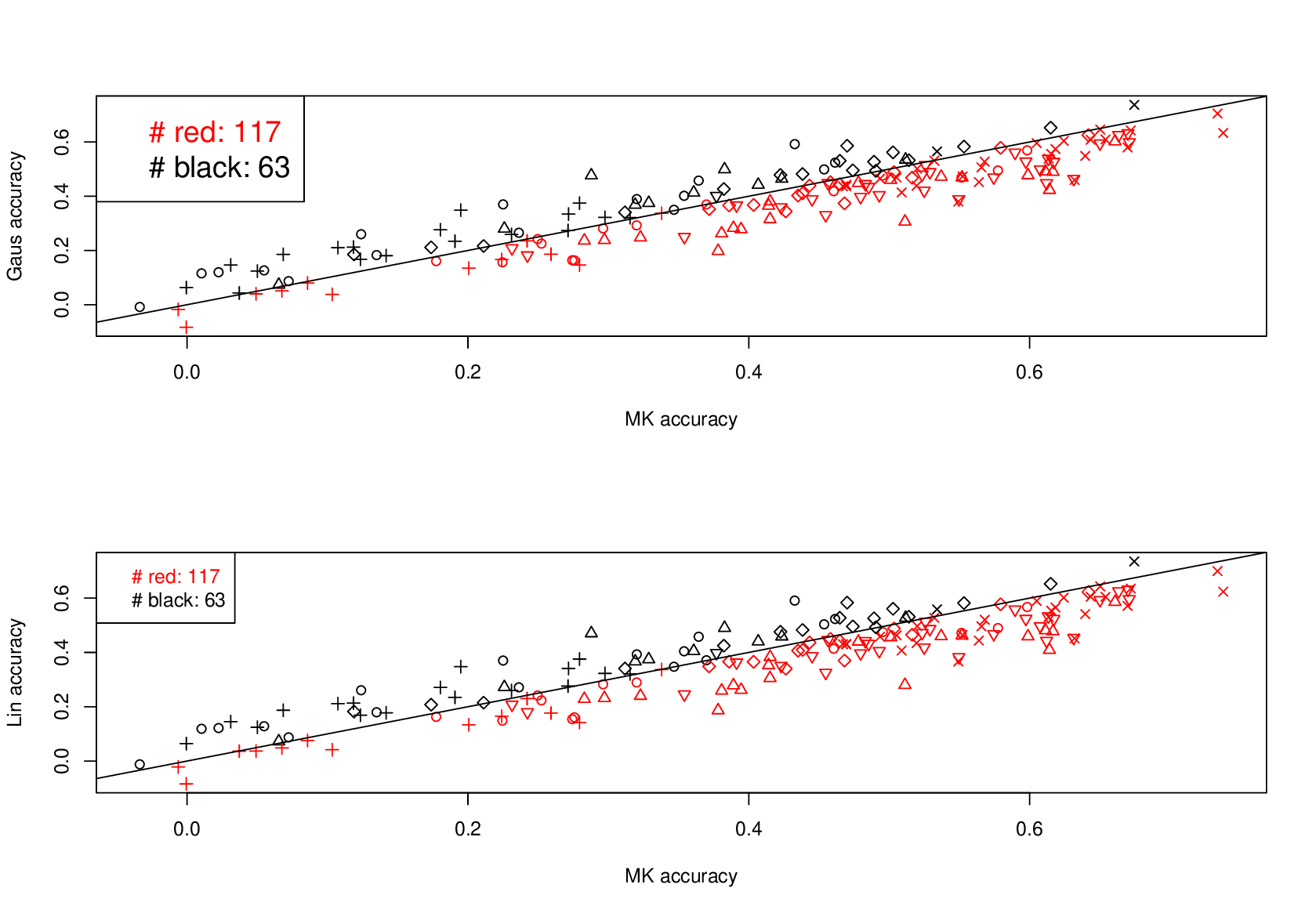}
\caption[Wheat Data: Accuracies 2]{Wheat Data: Accuracies of multiple kernel (MK) model (3 regions per chromosome) compared to Gaussian (Gaus) kernel model for the six traits. Red dots below the line correspond to cases MK model is more accurate than Gaus model. }
\label{fig:minipage49ex1}
\end{minipage}
\begin{minipage}[b]{0.45\linewidth}
\includegraphics*[angle=630,width=1\textwidth]{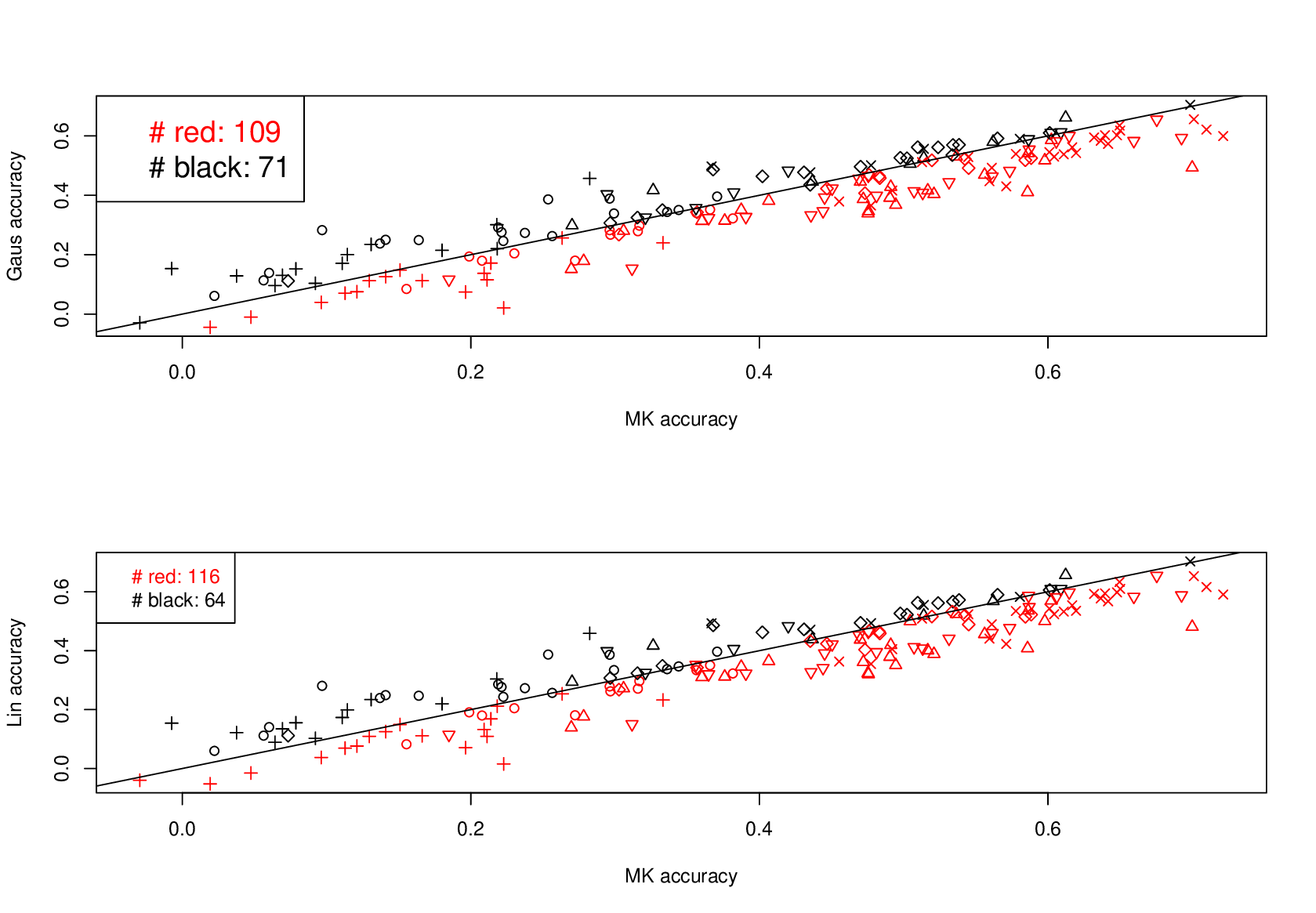}
\caption[Wheat Data: Accuracies 3]{Wheat Data: Accuracies of multiple kernel (MK) model (4 regions per chromosome) compared to Gaussian (Gaus) kernel model for the six traits. Red dots below the line correspond to cases MK model is more accurate than Gaus model. }
\label{fig:minipage79ex1}
\end{minipage}
\quad
\begin{minipage}[b]{0.45\linewidth}
\includegraphics*[angle=630,width=1\textwidth]{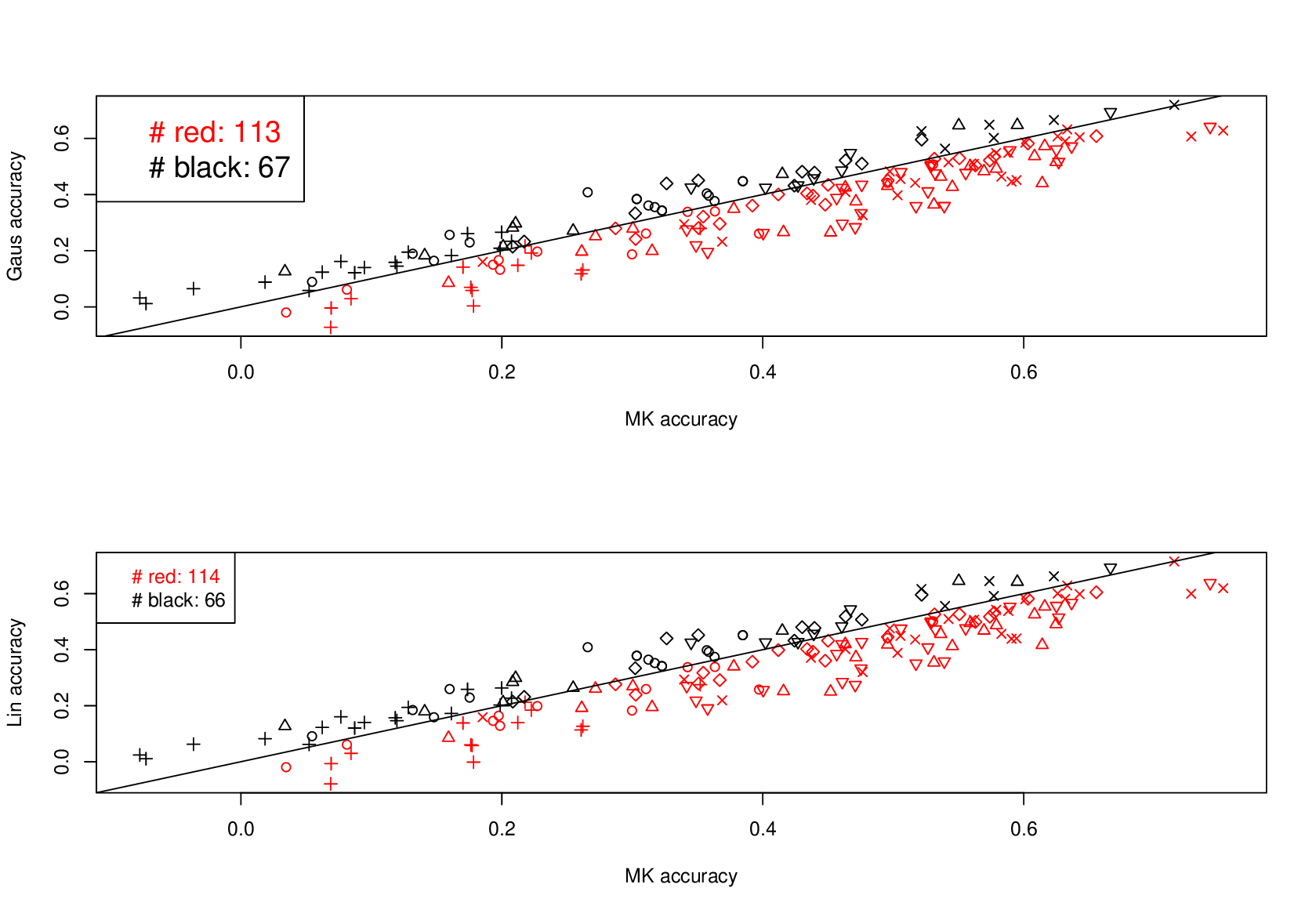}
\caption[Wheat Data: Accuracies 4]{Wheat Data: Accuracies of multiple kernel (MK) model (5 regions per chromosome) compared to Gaussian (Gaus) kernel model for the six traits. Red dots below the line correspond to cases MK model is more accurate than Gaus model. }
\label{fig:minipage89ex1}
\end{minipage}
\end{figure}

\begin{figure}[h]
\centering
\begin{minipage}[b]{0.45\linewidth}
\includegraphics*[angle=0,width=1\textwidth]{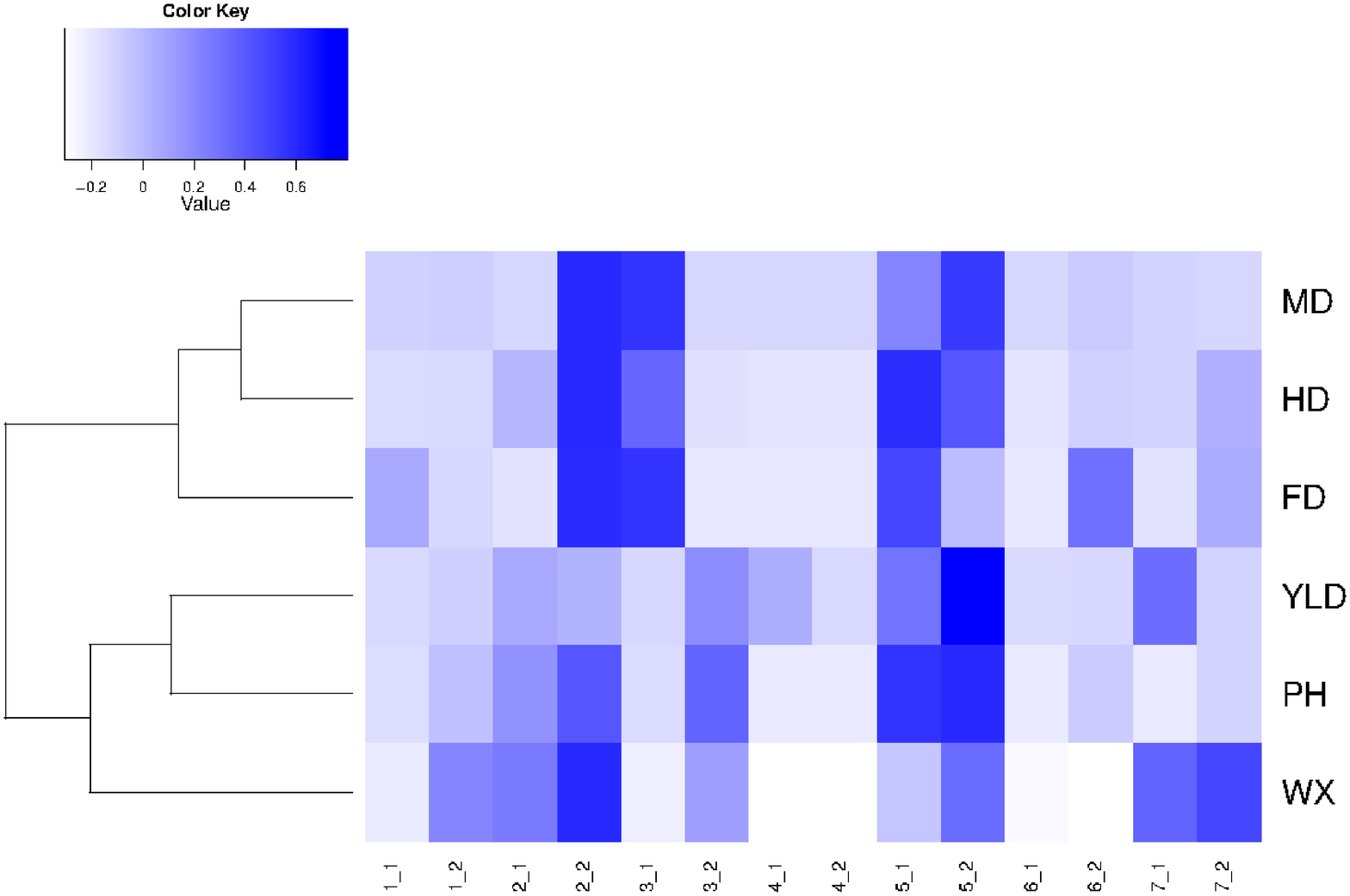}
\caption[Wheat Data: Associations 1]{Wheat Data: Associations from themultiple kernel (MK) model (2 regions per chromosome) for the six traits.}
\label{fig:minipage38ex1}
\end{minipage}
\quad
\begin{minipage}[b]{0.45\linewidth}
\includegraphics*[angle=0,width=1\textwidth]{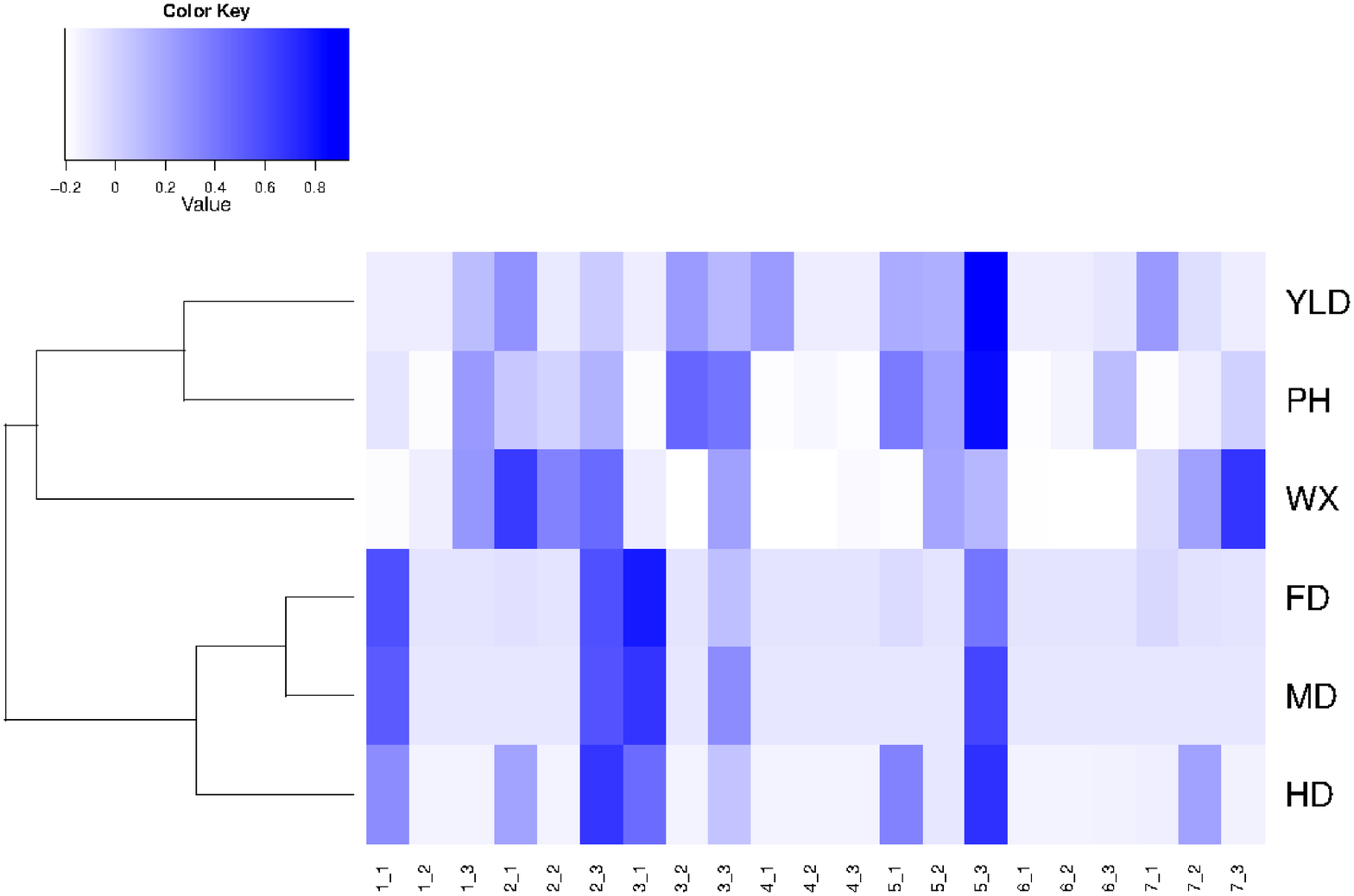}
\caption[Wheat Data: Associations 2]{Wheat Data: Associations from themultiple kernel (MK) model (3 regions per chromosome) for the six traits.}
\label{fig:minipage48ex1}
\end{minipage}
\begin{minipage}[b]{0.45\linewidth}
\includegraphics*[angle=0,width=1\textwidth]{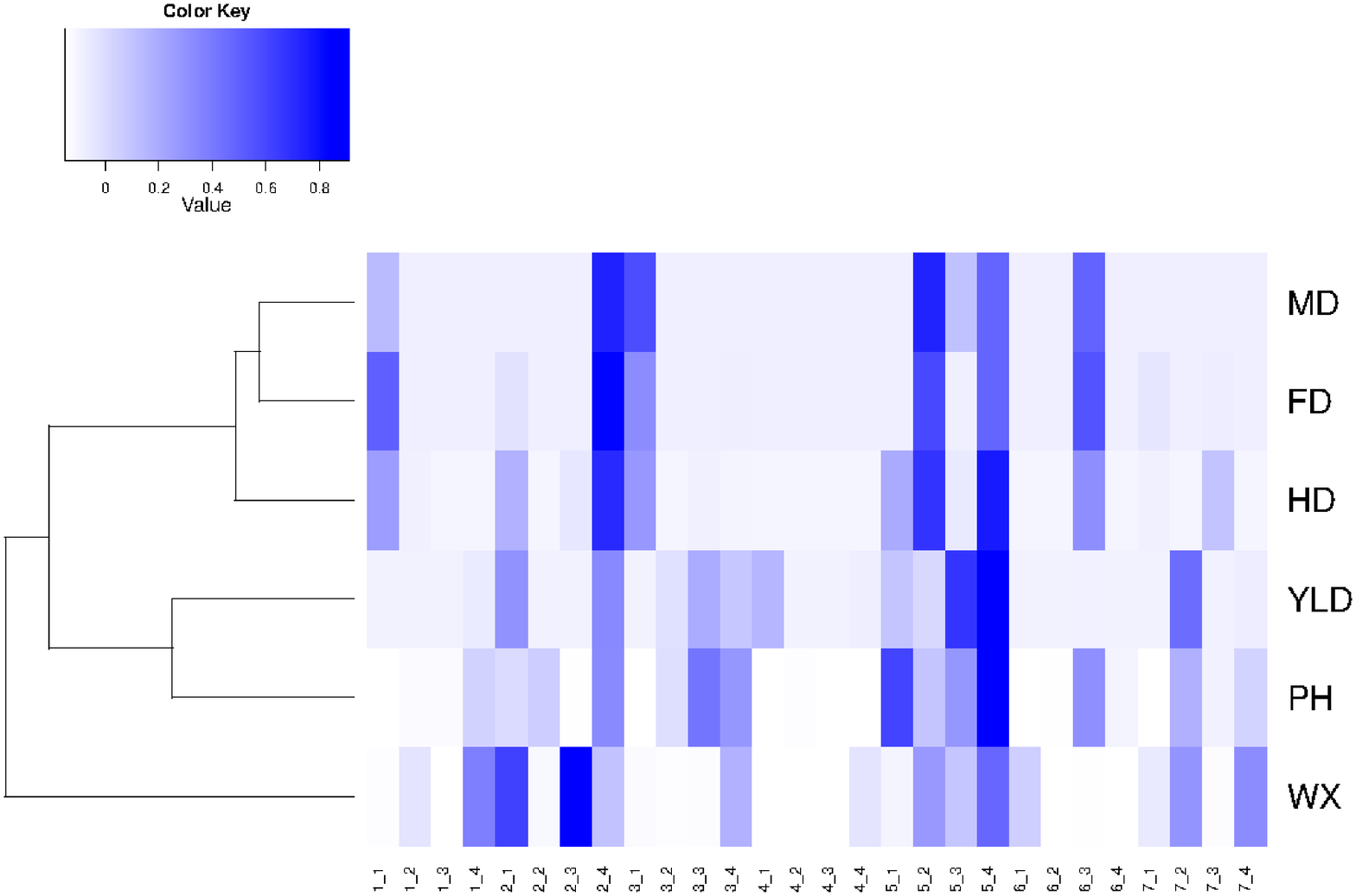}
\caption[Wheat Data: Associations 3]{Wheat Data: Associations from the multiple kernel (MK) model (4 regions per chromosome) for the six traits.}
\label{fig:minipage78ex1}
\end{minipage}
\quad
\begin{minipage}[b]{0.45\linewidth}
\includegraphics*[angle=0,width=1\textwidth]{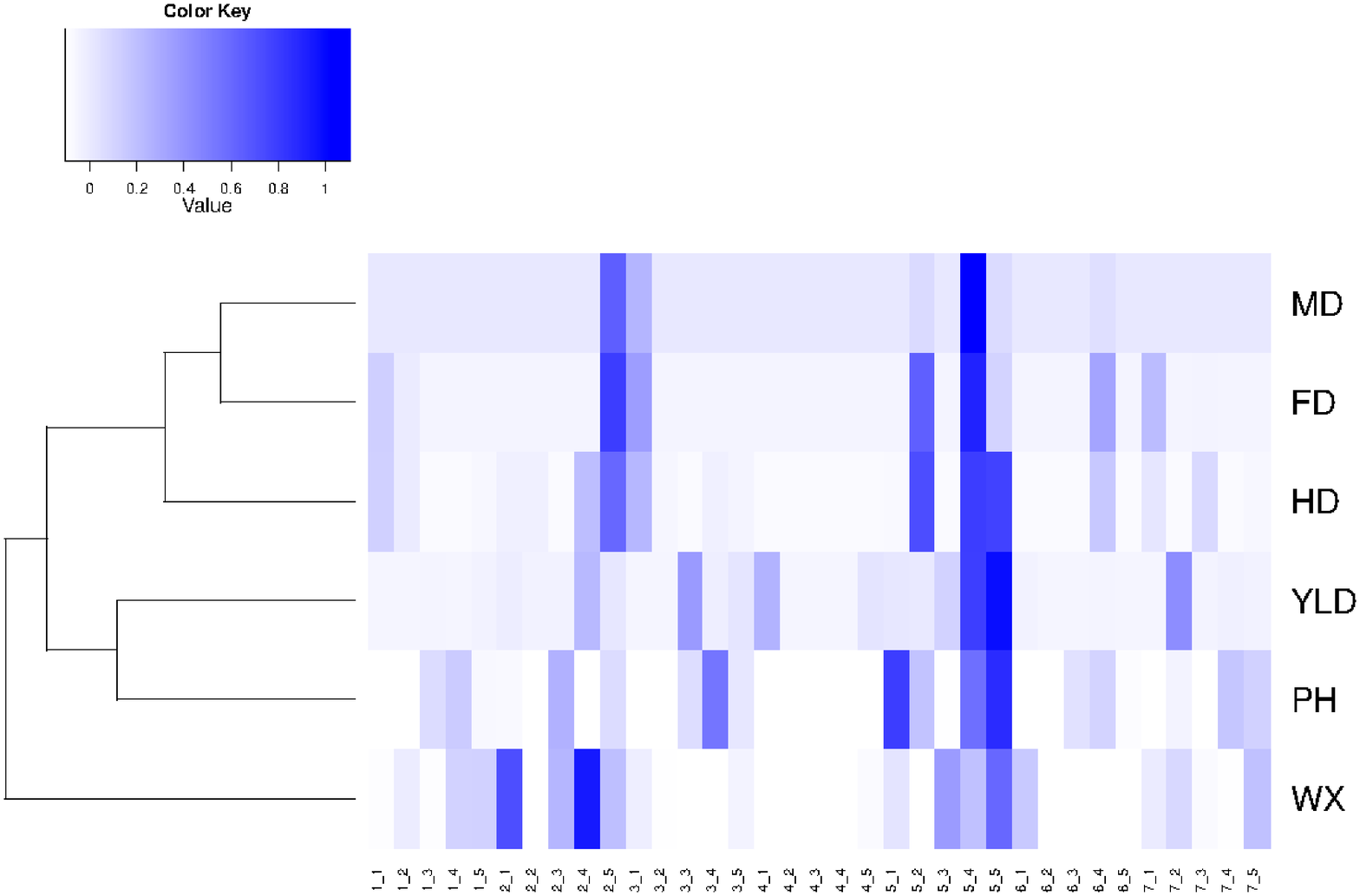}
\caption[Wheat Data: Associations 4]{Wheat Data: Associations from the multiple kernel (MK) model (5 regions per chromosome) for the six traits.}
\label{fig:minipage88ex1}
\end{minipage}
\end{figure}
\end{ex}

\begin{ex} (Mice Data)
The Mice data set we use for this analysis is available as a part of the R package SynbreedData \cite{wimmer2012package}.  Genotypic data consists of 12545 biallelic SNP markers and is available for 1940 individuals. The body weight at age of 6 weeks [g] and growth slope between 6 and 10 weeks age [g/day] are the measured for most of the individuals. The data was described in Solberg et al. \cite{valdar2006genome} and the heritabilities of these two traits are reported as 0.74 and 0.30 in Valdar et al \cite{valdar2006genome}. Here, we present the results from replication of the following experiment for the two traits at two different settings 30 times. A random sample of 1500 lines were selected in the training sample, a multiple kernel model (\ref{eq:additivemodel}), a Gaussian kernel model and a linear kernel model were trained and used to predict the genetic value of the individuals in the test data set, the accuracy defined as the correlation between the observed phenotypes in the test set and the corresponding as the estimated genotypic values are calculated for each model. The whole genome was divided in a similar fashion as displayed in Figure \ref{fig:fighthier} with a depth of 3 and only the regions in the most detailed level are used for multiple kernel model building. Two different models were obtained by using number of splits two and three at each level following the chromosomes (i.e., each chromosome was divided into 4 or 9 regions). The box-plots comparing accuracies of the models and the mean importance scores for different regions from the multiple kernel models over 30 replications are displayed in Figures \ref{fig:minipage3ex1}, \ref{fig:minipage4ex1}, \ref{fig:minipage7ex1} and \ref{fig:minipage8ex1} correspondingly. For both traits the multiple kernel model is substantially more accurate. In addition, the the association derived as an output to this model supports the previously reported association of body weight related traits to the X chromosome \cite{dragani1995mapping}.  

\begin{figure}[h]
\centering
\begin{minipage}[b]{0.45\linewidth}
\includegraphics*[angle=0,width=1\textwidth]{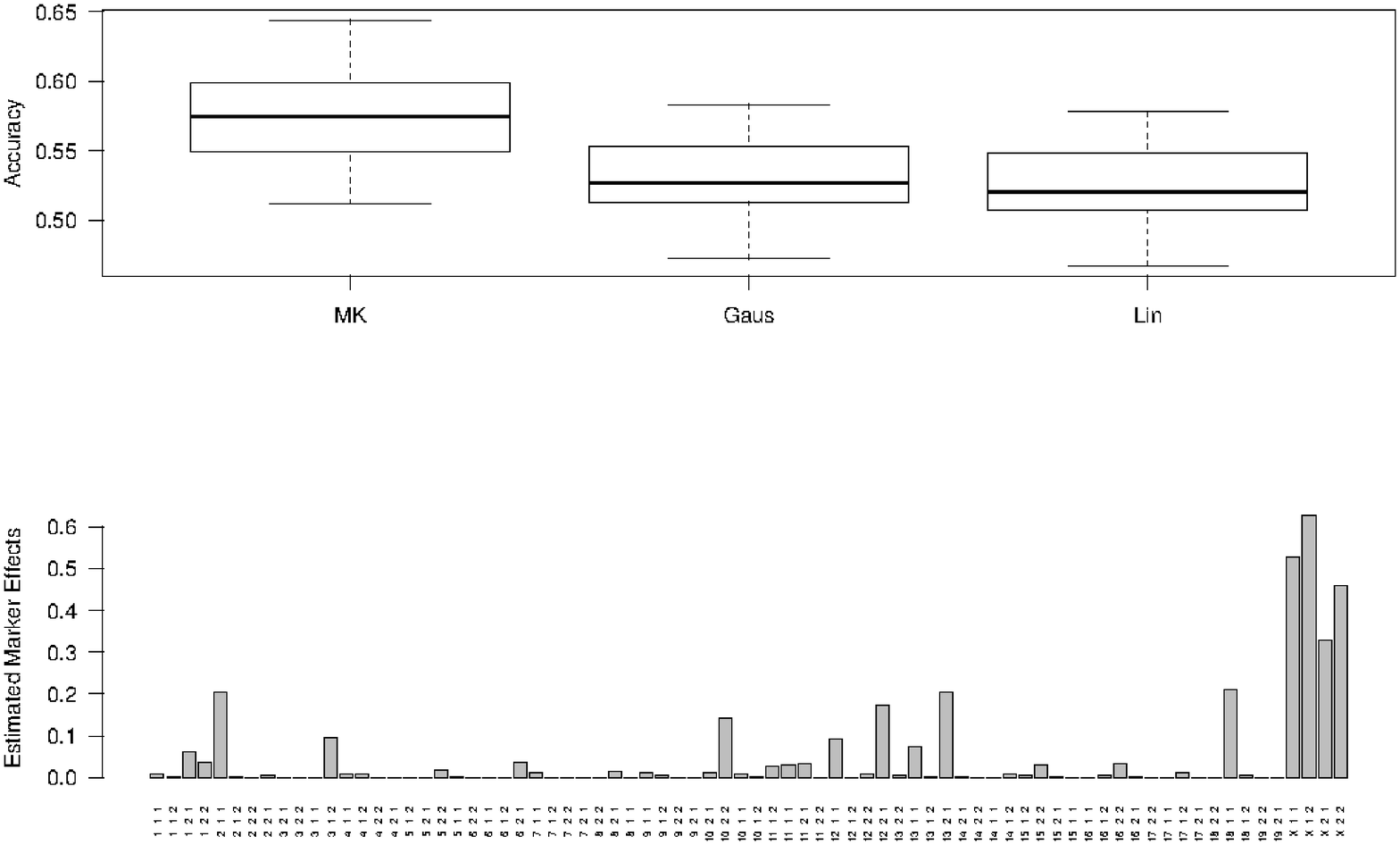}
\caption[Mice Data: Accuracies and Associations 4]{Mice Data: Accuracies and associations for multiple kernel (MK) (4 regions per chromosome) and accuracies for linear kernel (Lin) and Gaussian kernel (Gaus) models  for ''weight at age of 6 weeks [g]''.}
\label{fig:minipage3ex1}
\end{minipage}
\quad
\begin{minipage}[b]{0.45\linewidth}
\includegraphics*[angle=0,width=1\textwidth]{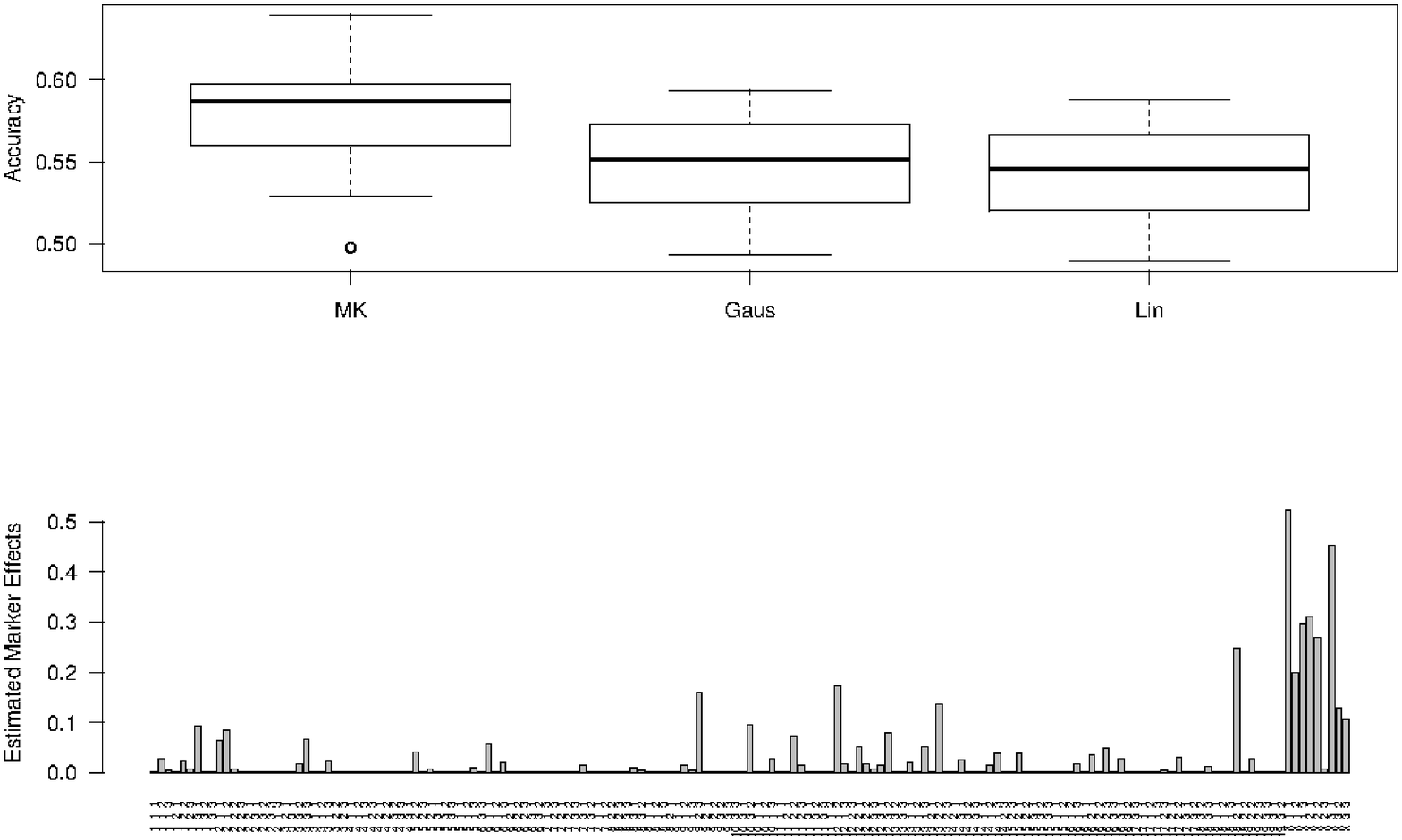}
\caption[Mice Data: Accuracies and Associations 4]{Mice Data: Accuracies and associations for multiple kernel (MK) (9 regions per chromosome) and accuracies for linear kernel (Lin) and Gaussian kernel (Gaus) models for ''weight at age of 6 weeks [g]''.}
\label{fig:minipage4ex1}
\end{minipage}
\begin{minipage}[b]{0.45\linewidth}
\includegraphics*[angle=0,width=1\textwidth]{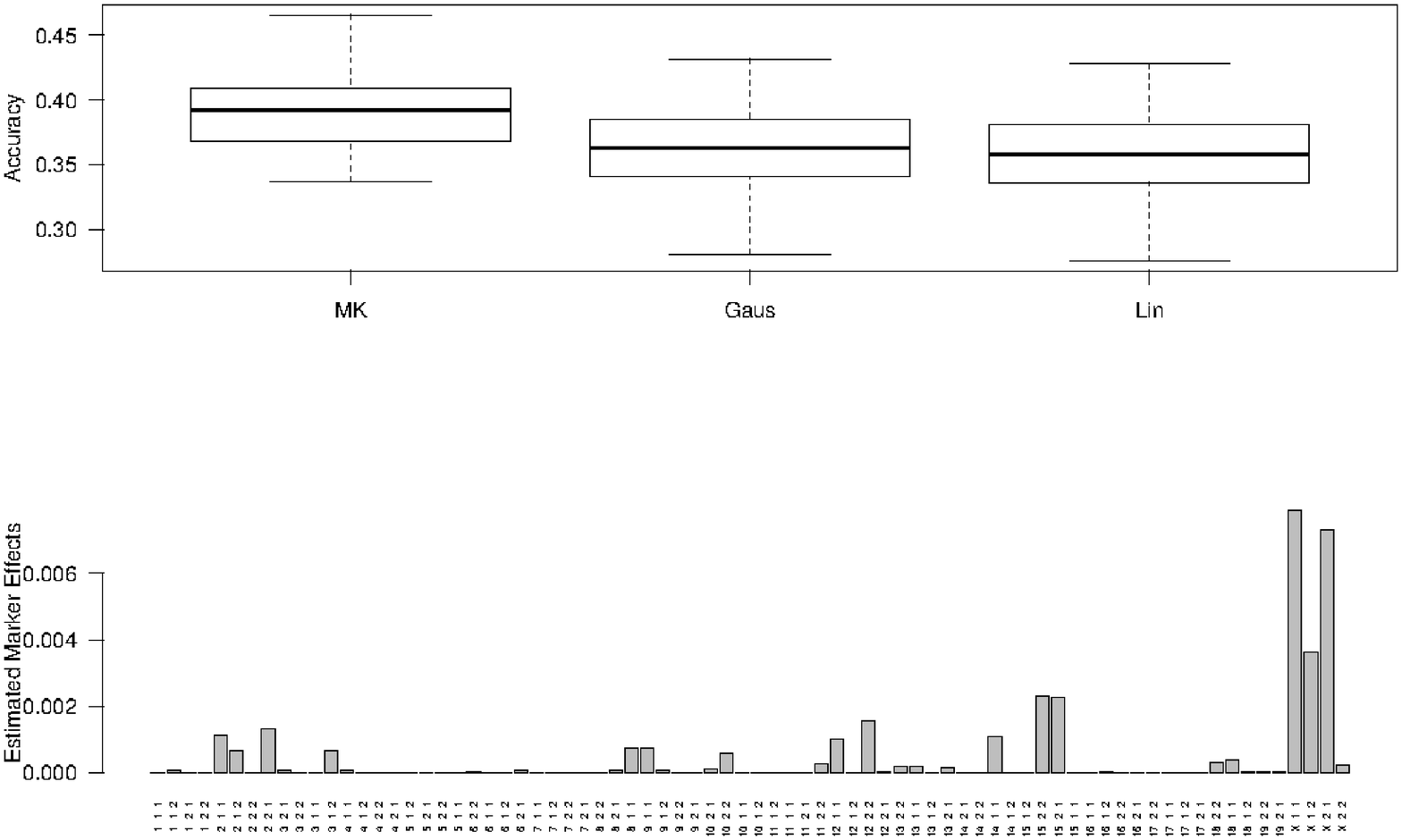}
\caption[Mice Data: Accuracies and Associations 4]{Mice Data: Accuracies and associations for multiple kernel (MK) (4 regions per chromosome) and accuracies for linear kernel (Lin) and Gaussian kernel (Gaus) models  for ''growth slope between 6 and 10 weeks age [g per day]''.}
\label{fig:minipage7ex1}
\end{minipage}
\quad
\begin{minipage}[b]{0.45\linewidth}
\includegraphics*[angle=0,width=1\textwidth]{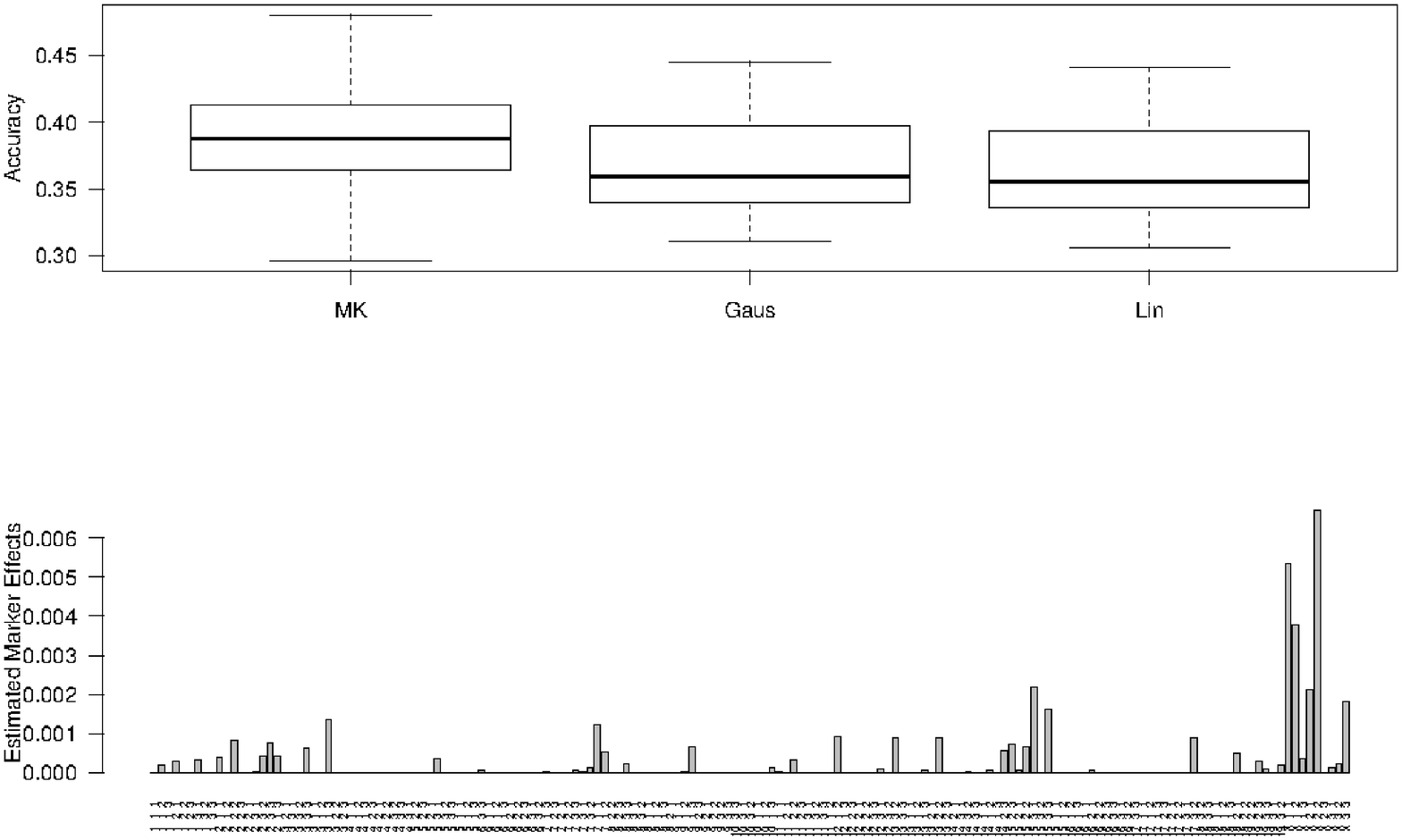}
\caption[Mice Data: Accuracies and Associations 4]{Mice Data: Accuracies and associations for multiple kernel (MK) (9 regions per chromosome) and accuracies for linear kernel (Lin) and Gaussian kernel (Gaus) models  for ''growth slope between 6 and 10 weeks age [g per day]''.}
\label{fig:minipage8ex1}
\end{minipage}
\end{figure}
\end{ex}

\begin{ex}(Barley Data) Tocotrienols are class of fat soluble chemical compounds related to vitamin E activity.  Vitamin E deficiency is connected to many health problems therefore increased levels of these compounds is a desirable property for crops. In an experiment carried out by USDA-ARS  during the years 2006-2007, tocotrienol levels for 1723 barley lines were recorded in total of 4 environments (2 years and 3 locations).  2114 markers on 7 chromosomes were available for the analysis. The whole genome was divided in a similar fashion as displayed in Figure \ref{fig:fighthier}.  We have sampled 1500 lines for training the models and we have used the rest of the lines to  evaluate the fit of our models. The whole genome was divided in a similar fashion as displayed in Figure \ref{fig:fighthier} with a depth of 3 withs 2 splits at each level and only the regions in the most detailed level are used for multiple kernel model building. Accuracies and associations

\begin{figure}[h]
	\centering
\includegraphics*[width=1\textwidth]{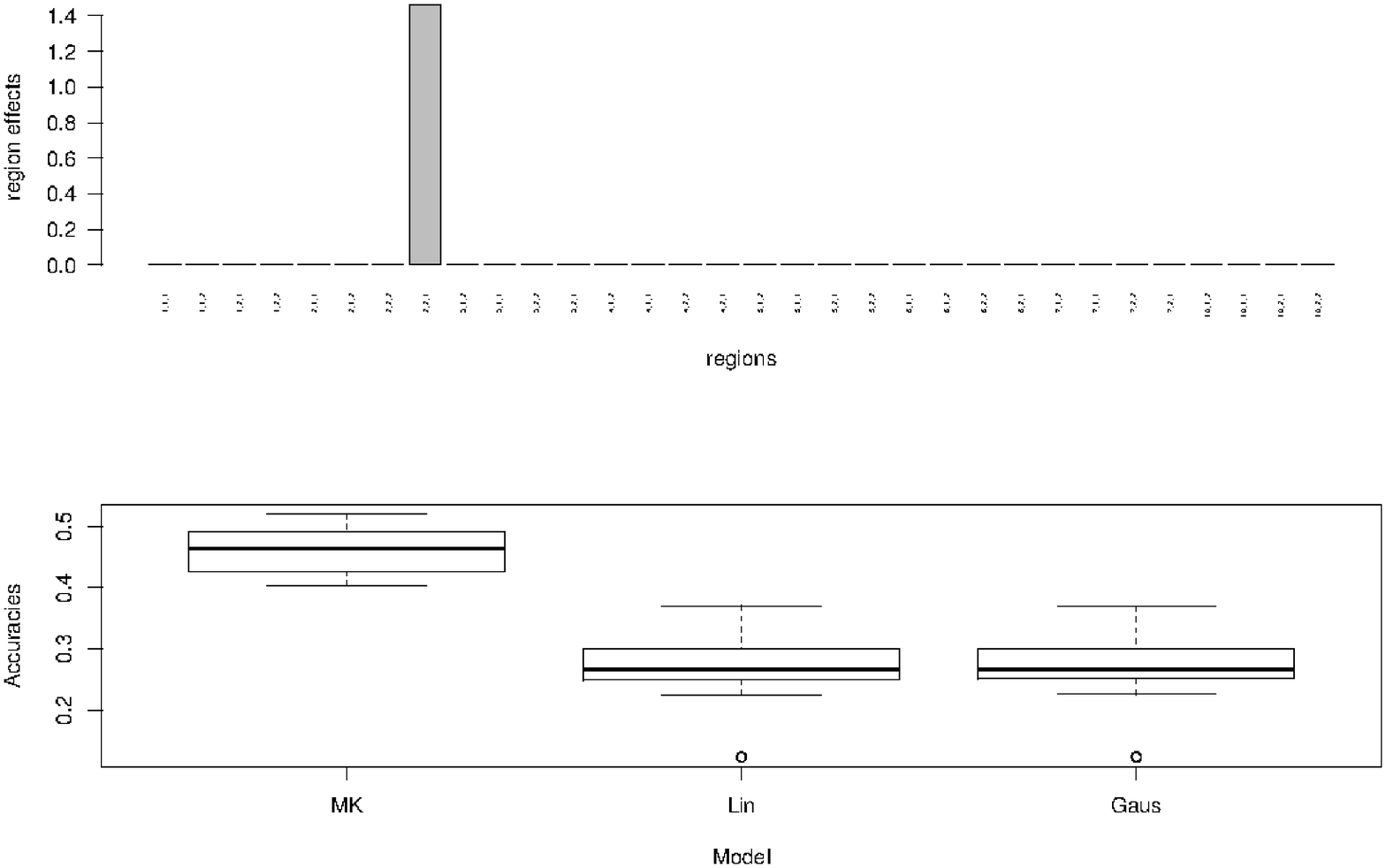}
	\caption[Barley Data: Accuracies and Associations]{Maize Data: Accuracies and associations for multiple kernel (MK) (4 regions per chromosome) and accuracies linear kernel (Lin) and Gaussian kernel (Gaus) models for tocotrienol levels.}
	\label{fig:barleydata3}
\end{figure}

\end{ex}

\begin{ex}(Rice Data) A diverse collection of 395 O. sativa (rice) accessions including both landraces and elite varieties which represent the range of geographic and  genetic diversity of the species was used in this example. In addition to measurements for 36 continuous traits, the 40K SNP targets were available for these 395 accessions.  All of the data from this study are publicly available at  \url{www.ricediversity.org}. This data was first presented in \cite{zhao2011genome} and was also analyzed in \cite{wimmer2013genome}. We have used the SPMM's with Gaussian (Gaus), linear (lin) and the locally epistatic model with lasso postprocessing for the all the traits. The locally epistatic model included effects of whole genome and whole chromosomes in addition to five equal length sections within each chromosome. The importance scores for the multiple kernel model for the 36 traits are displayed in  \ref{fig:ricedata2}, the similarities of the traits based on the importance scores of the regions are summarised by the dendogram on the vertical axis. The correlation of the estimated genomic component and the trait values for the individuals in the test data set ($20\%$ of the 395 individuals selected at random for which) are calculated for $30$ independent instances and the results are summarized  in Figure \ref{fig:ricedata3}.

\begin{figure}[h]
	\centering
\includegraphics*[angle=270,width=1\textwidth]{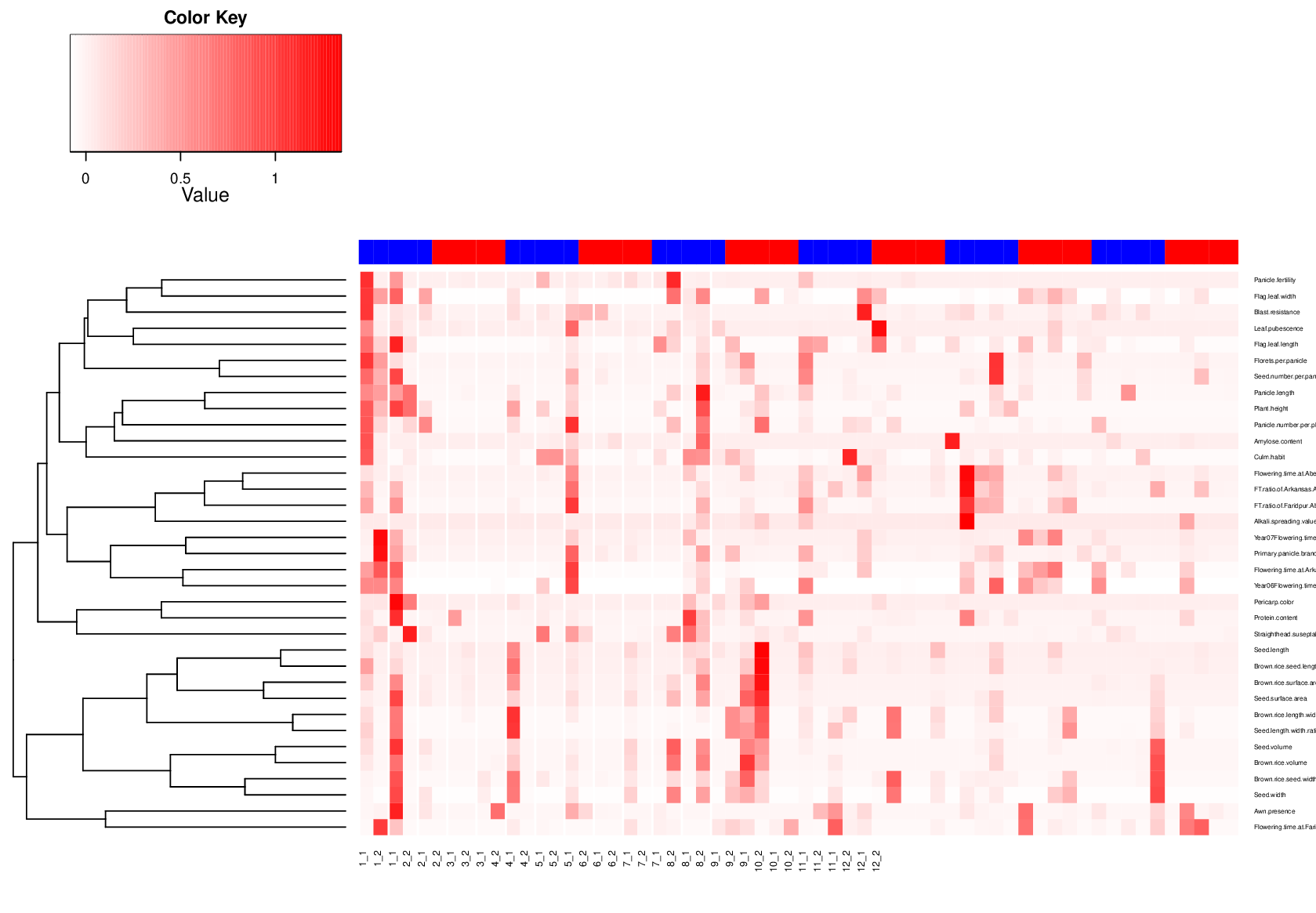}
	\caption[Rice Data: Associations]{Rice Data: Associations from the multiple kernel (MK) model (5 regions per chromosome) for the 36  traits.}
	\label{fig:ricedata2}
\end{figure}

\begin{figure}[h]
	\centering
\includegraphics*[angle=270,width=1\textwidth]{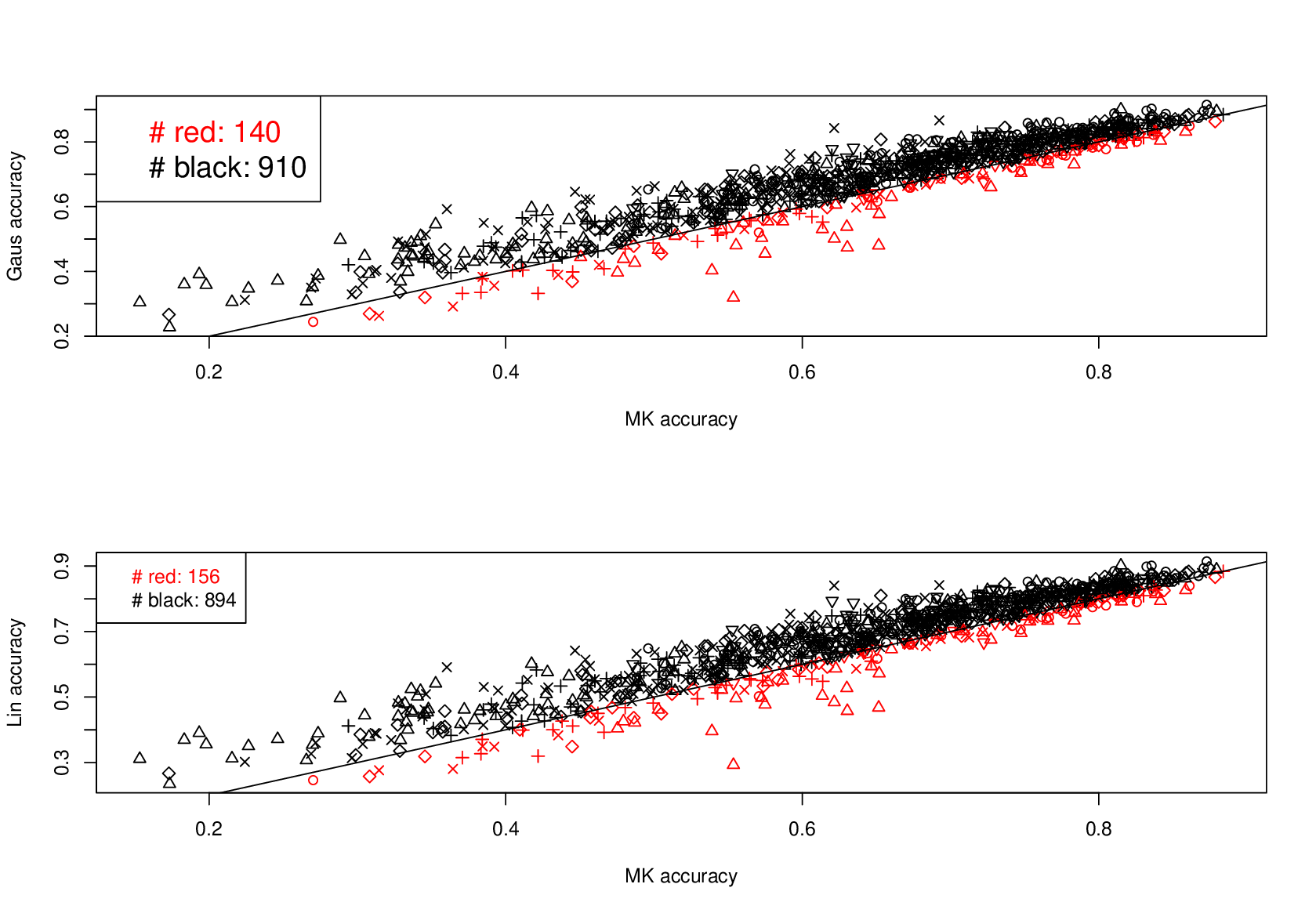}
	\caption[Rice Data: Accuracies]{Rice Data: Accuracies of multiple kernel (MK) model (5 regions per chromosome) compared to Gaussian (Gaus) kernel model for the 36 traits.}
	\label{fig:ricedata3}
\end{figure}

\end{ex}

\begin{ex}(Maize Data) 
This data is given in \cite{romay2013comprehensive} and was also analysed in \cite{wimmer2013genome}. 68120 markers on 2279 USA national inbred maize lines and their phenotypic means for degree days to silking compose the data set. Accuracies for multiple kernel (MK) (5 regions per chromosome), linear kernel (Lin) and Gaussian kernel (Gaus) models for degree days to silking and the imporance scores from the MK model are displayed in  Figure \ref{fig:maizedata} .

\begin{figure}[h]
	\centering
\includegraphics*[angle=0,width=1\textwidth]{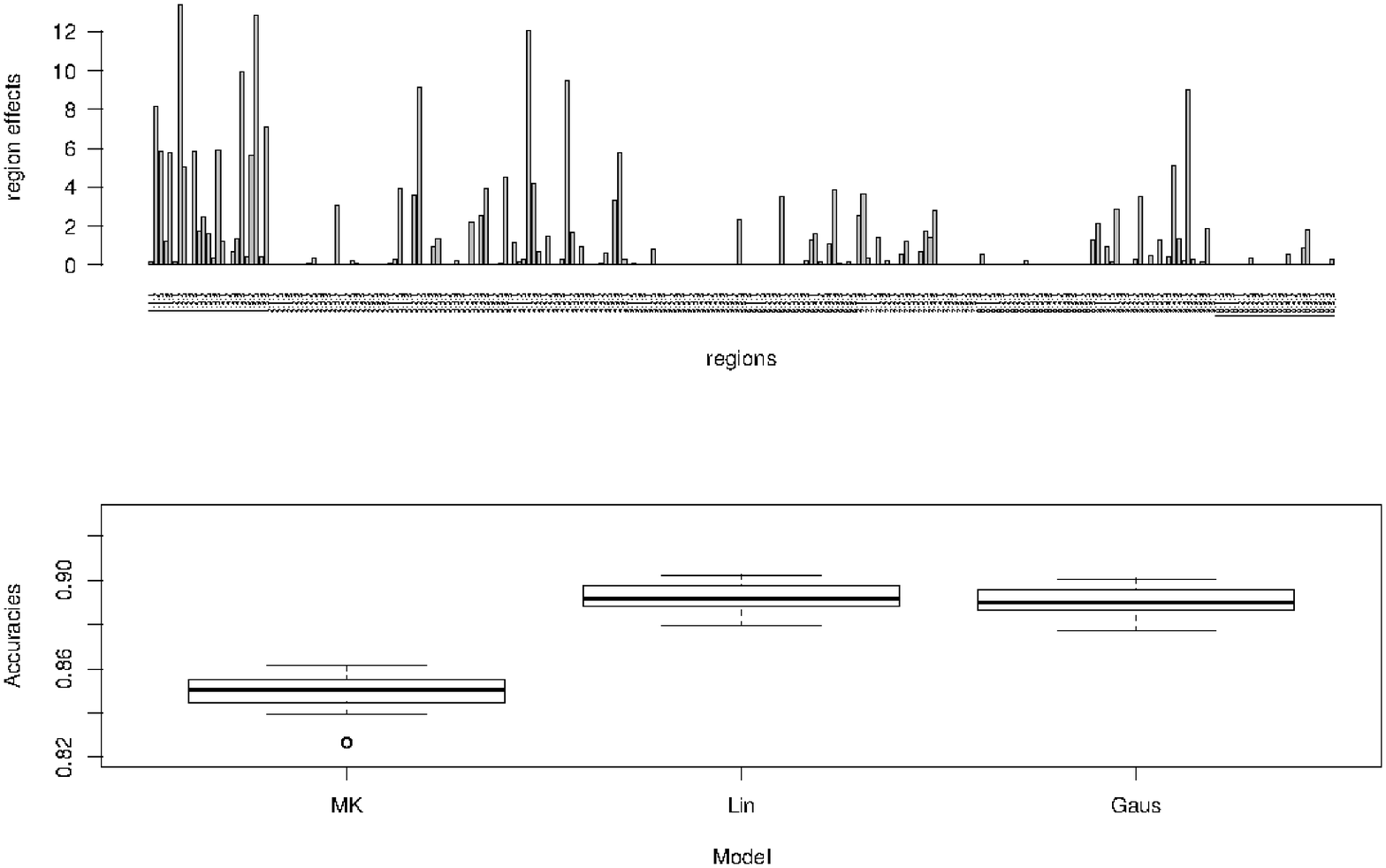}
	\caption[Maize Data: Accuracies and Associations]{Maize Data: Accuracies and associations for multiple kernel (MK) (25 regions per chromosome) and accuracies for linear kernel (Lin) and Gaussian kernel (Gaus) models for degree days to silking.}
	\label{fig:maizedata}
\end{figure}

\begin{figure}[h]
	\centering
\includegraphics*[angle=0,width=1\textwidth]{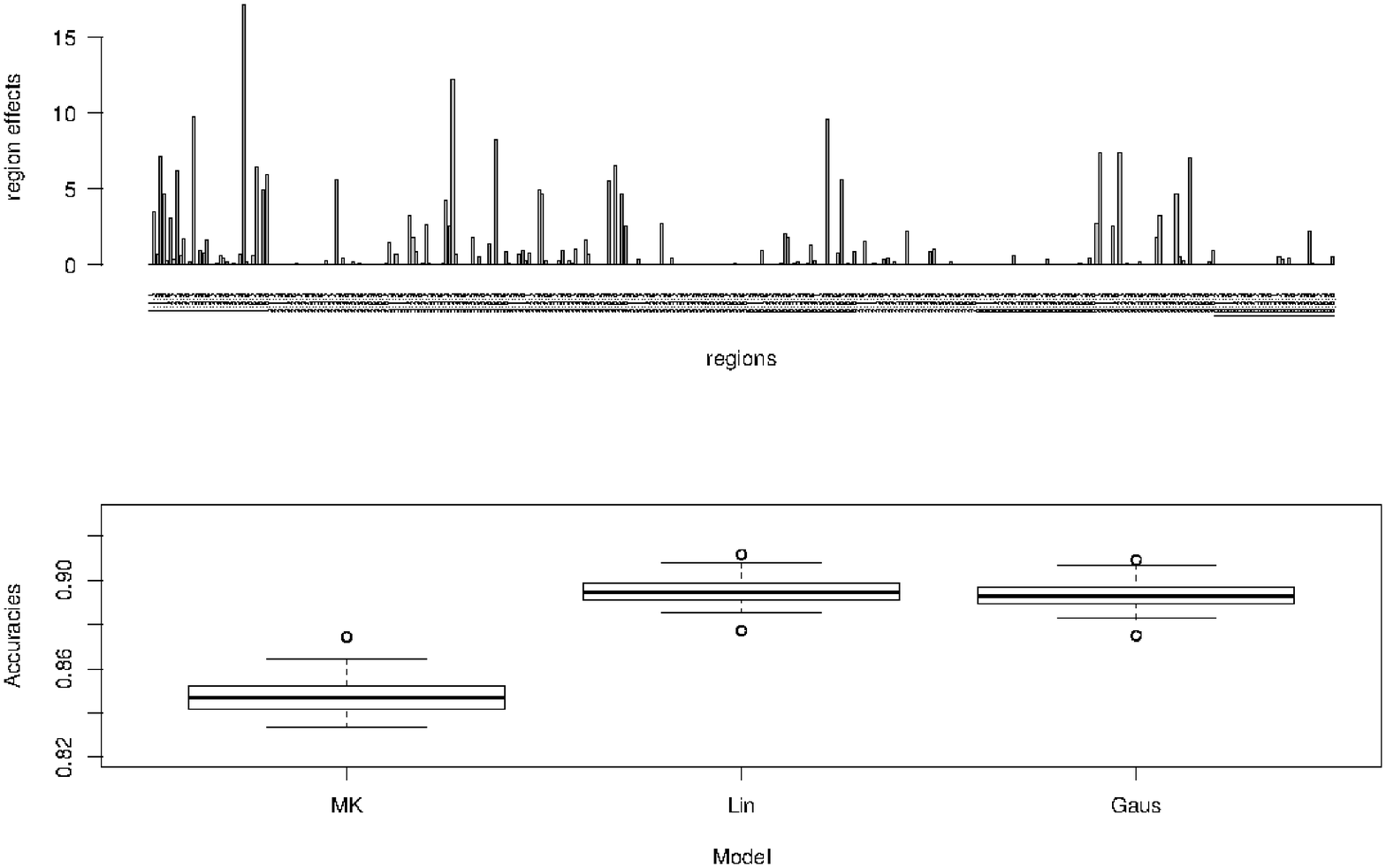}
	\caption[Maize Data: Accuracies and Associations]{Maize Data: Accuracies for multiple kernel (MK) (36 regions per chromosome), linear kernel (Lin) and Gaussian kernel (Gaus) models for degree days to silking.}
	\label{fig:maizedata}
\end{figure}

\end{ex}

\section{Conclusions}

The multiple kernel models proposed in this paper have good accuracy and explanatory value. Although it seems to depend on complexity of the trait / population structure, similar or better accuracies were obtained for a number of populations compared to single kernel models. The multiple kernel models have the additional advantage that only a small fraction of genomic regions are utilized in the final model and the importance scores for these regions are readily available as an output to the model.

The approaches introduced allows us to use the markers in naturally occurring blocks. In the context of the SPMM in (1) there are very fast algorithms that can take advantage of this dimension reduction. For the linear kernel function, the order of calculations to solve a SPMM with one kernel matrix is proportional to $min(n,m)$ where $m$ here is the number of features in that kernel. No matter what the input dimension is SPMM parameter estimation involves matrices of order $n.$   Therefore, the multiple kernel approach overcomes the memory problems that we might incur when the number of markers is very large by loading only subsets of markers in the memory at a time.

The local kernels use information collected over a region in the genome and, because of linkage, will not be effected by a few missing or erroneous data points, so this approach is also robust to missing data and outliers.  

As mentioned earlier, we can obtain local kernel matrices by defining regions in the genome and calculating a separate kernel matrix for each group and region. The regions can be overlapping or discrete. If the some markers are associated with each other in terms of linkage or function it might be useful to combine them together.  The whole genome can be divided physically into chromosomes, chromosome arms or linkage groups. Further divisions could be based on recombination hot-spots, or just merely based on local proximity. We could calculate a separate kernel for introns and exons, non coding, promoter or repressor sequences.  We can also use a grouping of markers based on their effects on low level traits like lipids, metabolites, gene expressions, or based on their allele frequencies. When some markers are missing for some individuals, we can calculate a kernel for the presence and absence states for these markers. When no such guide is present one can use a hierarchical clustering of the variables. It is even possible to incorporate group memberships probabilities for markers so the markers have varying weights in different groups. We intend to address these and some other related issues in subsequent work.

\section{Supplementary Materials}
\begin{description}

\item[Title:] Data sets and R-codes used for the illustrations. (MultipleKernel.tar) (GNU zipped tar file)

\end{description}

\section*{Acknowledgments}
This research was supported by the USDA-NIFA-AFRI Triticeae Coordinated Agricultural Project, award number 2011-68002-30029.

\bibliographystyle{plain}

\bibliography{kernelbib}

\end{document}